\documentstyle[12pt]{article}

\begin{document}
\title{Non-Gaussian Distributions in Extended Dynamical Systems}
\author{Ravi Bhagavatula and C. Jayaprakash \\
 Department of Physics \\
 The Ohio State University \\}
\maketitle
\begin{abstract}

We propose a novel mechanism for the
origin of non-Gaussian tails in the probability distribution
functions (PDFs) of local variables in nonlinear, diffusive,
dynamical systems including passive scalars advected by chaotic
velocity fields. Intermittent
fluctuations on appropriate time scales
in the amplitude of the (chaotic) noise
can lead to exponential tails. We provide
numerical evidence for such behavior in
deterministic, discrete-time passive scalar models.
Different possibilities for PDFs
are also outlined.
\end{abstract}


\vspace{0.1in}
\pagebreak

 The explanation of the form  of the
probability distribution of the fluctuations in local variables
in extended, dissipative, dynamical
systems poses an interesting challenge.
In various turbulence experiments \cite{cetal,jw,gol,cgh},
Probability Distribution Functions (PDFs)
of velocity gradients and passive scalars (temperature)
are observed to be non-Gaussian; in particular,
exponential tails have been seen whereas Gaussian
distributions might be expected if one naively
invokes the central limit theorem.
Theoretically, a variety of explanations including an ingenious
phenomenological model based on a nonlinear
Fokker-Planck equation \cite{pss} and methods
that rely on obtaining closure
and moment balance relations \cite{sy,ycck} have been proposed
to explain exponential tails in the passive scalar problem.
The moment balance approach has been used to provide good fits
to experimental data \cite{ching}. In addition, several eddy diffusive
models that yield exponential tails have
also been studied \cite{ker}.

In this letter we explore a new mechanism in which the nature of the
temporal correlations of the fluctuations that couple to a diffusive field
plays a crucial role. We focus on passive scalar models
and show that non-Gaussian PDFs
including {\em exponential} and  stretched
exponential behaviors can arise when local
variables are coupled to fluctuations whose amplitude varies
randomly on a time scale comparable to the
intrinsic diffusive time scale.
In experimental fluid systems, exponential tails
could, therefore, arise if the velocity fluctuations
exhibit {\em intermittent} behavior in their magnitude on relevant time
scales.

We consider the diffusion of the temperature
field $\theta (\vec r, t)$
advected by a flow characterized by the
velocity $\vec V$ in the presence of an
externally-imposed  mean temperature gradient $\beta$.
This is described by the passive scalar equation
\begin{equation}
\frac{\partial \theta}{\partial t} = \nu \nabla^2 \theta
 - \vec{V} \cdot \vec{\nabla} \theta.
\end{equation}
The coefficient $\nu$ is the effective thermal diffusivity
and the velocity is assumed to be
incompressible: $\vec {\nabla} \cdot \vec{V}=0$.

 The behavior of (1) in the presence of prescribed random velocity fields
can be studied numerically directly in the continuum. However, the
case of a velocity field arising from
a deterministic chaotic dynamics is computationally prohibitive if one
uses Navier-Stokes equations. Instead we consider 2-dimensional
coupled map lattice models in discrete time and space which can be
viewed as coarse-grained versions of the continuum
equation with the lattice spacing of the order of the
correlation length of the velocity fluctuations.
The scalar $\theta (i)$ is defined on each site $i$
of a $L \times L$ square lattice with
a mean gradient $\beta$ along the $y$ direction
and periodic boundary conditions in the $x$ direction.
The system evolves synchronously in discrete time $n$ according to
\begin{equation}
\theta_{n+1}(i)=\theta_{n}(i)+\nu \nabla _L^2 \theta_n(i)
- \vec V_n(i) \cdot \vec {\nabla}_L \theta_n(i)
\end{equation}
where $\vec \nabla _L$ is a symmetric, lattice gradient.
The incompressibility of the two-dimensional
velocity field $\vec V_n(i) = (u_n(i),v_n(i))$ is enforced by
obtaining it from a stream function $\psi_n(i)$:
$u_{n}(i) = \gamma \nabla_y \psi_n(i);
 v_{n}(i) = -\gamma \nabla_x \psi_n(i)$.
The velocity vanishes at both $y=1$ and $y=L$ boundaries.
The parameter $\gamma$ is introduced to adjust
the variance $\sigma_v$ of the velocity field
since the (discrete time) model is unstable
for large values of the variances \cite{bound}.

 In our numerical simulations, for all the models we explored,
the mean scalar profile is linear. If we expand $\theta$ around
the mean profile we obtain both an {\em additive}
noise term, $\beta v_n(i)$, and a
convective ({\em multiplicative}) noise term.
This separation also occurs in the
continuum equation. We find that different regimes of behavior
of the scalar PDFs can be characterized empirically by a single parameter
$B$ that measures the relative strengths of
convective and additive noise terms: $B=\sigma_{\theta}/\beta\xi_{\theta}$,
where $\sigma_{\theta}$ is the variance and
$\xi_{\theta}$ the characteristic length scale of $\theta$.

We first consider the regime $B < 1$ which
occurs for  $\sigma_v^2\tau_c/\nu < 1$
where $\tau_c$ is the correlation time of the velocity field.
Different models are defined by the dynamics of $V_n(i)$.
{\em When the velocity field has Gaussian amplitude fluctuations
on a time scale $\tau_a$ comparable to the diffusive time scale
$\tau_d = 1/\nu$ we obtain PDFs with exponential tails for $\theta$}.

Model A: The easiest way to obtain amplitude fluctuations is to use a
{\em stochastic} model and choose
the stream function to be a product of two noise terms
$\eta_n(i)\eta'_n(i)$ where
$\eta'_n(i)$ is a white noise and $\eta_n(i)$ a Gaussian noise with a
correlation time $\tau_a$, both having delta-function
spatial correlations. The stream function and
the velocity display two correlation times: (i)
the mixing time $\tau_c$ governed by $\eta'_n(i)$ chosen to be
one timestep and (ii) the time scale $\tau_a$ on which
the amplitude varies. This model exhibits exponential tails
in the PDF of $\theta(i)$ when
$\tau_a \approx \tau_d$ (See Fig.1). On the other hand
when $\tau_a << \tau_d$ the distributions are Gaussian.

Model B: We next provide numerical evidence to show that
exponential tails also occur when the stream function is
derived from a {\em deterministic} chaotic model provided it leads
to amplitude fluctuations on the appropriate time scale.
We investigate the following model:
define the stream function in terms of
an auxiliary variable
$\phi_n(i)$ by $\psi_n(i) = \nabla _L^2 \phi_{2n}(i)$; let
$\phi_n(i)$ be updated according to
$\phi_{m+1}(i)=\frac{1}{5} \sum_k F(\phi_m(k))$, where
the sum includes the site $i$ and
its nearest neighbours. The function $F$
is chosen to be $F(x)=(1-2a)x+2ax^3$;
we consider $a=2$ for which
the maximum liapunov exponent is $\approx 0.95$
and the correlation length is measured to be
one lattice spacing. The PDFs for $u,v$ display Gaussian
tails (See Fig.2a). The PDF for $\theta(i)$ is shown
in Fig.2b which clearly shows exponential
tails over five decades (similar results persist
for a range of values of $a$).
The slope of the tails is proportional to
$\sqrt {\nu}/(\beta\sigma_v)$ as expected from
dimensional considerations. To demonstrate that the velocity
field derived from this deterministic model
displays intermittent amplitude fluctuations
we compute the autocorrelation functions
$C_2(n)=<v_nv_0>-<v_0>^2$
and $C_4(n)=<v^2_nv^2_0>-<v^2_0>^2$,
where $<..>$ indicates time average.
In Fig.3 we show $C_4(n)$ which
shows two exponential decays
associated with time scales
$\tau_c /2 \approx 1$, $\tau_a/2 \approx 4$.
The inset shows $C_2(n)$ which is expected to display a
decay time $(1/\tau_c + 1/\tau_a)^{-1} \approx 1.6$,
consistent with the above interpretation.
Note that in Fig.2b $\tau_d$ is comparable to $\tau_a$
and the kurtosis decreases with increasing $\tau_d$.

 To elucidate this mechanism we present a
mean-field-like toy model with only additive noise since
the multiplicative noise term is small for $B < 1$.
The model is described by the Langevin equations,
\begin{eqnarray}
\dot{x} & = & -\nu x + y \eta_1,  \\
\dot{y} & = & -\alpha y + \eta_2,
\end{eqnarray}
 where the variable $x$ couples to a noise with an amplitude
$y$ that fluctuates in time. The terms $\eta_i$ are assumed to be
white noise with $<\eta_i>=0$ and $<\eta_i(t)\eta_j(t')>
=\sigma_i^2 \delta_{ij}\delta(t-t')$.
The crucial time scales are given by $\tau_a=1/\alpha$ and $\tau_d=1/\nu$.
It is convenient to introduce scaled variables $x_0=x/\sqrt{D_1D_2}$,
and $y_0=y/\sqrt{D_2}$ where $D_1 = \sigma_1^2/2\nu$,
$D_2 = \sigma_2^2/2\alpha$.
Since $y_0$ is independent
of $x_0$ one can obtain
a stationary solution for the Fokker-Planck
(FP) equation \cite{risk} of the form
$P(x_0,y_0)=Q(x_0|y_0)R(y_0)$,
where $R(y_0) = [2\pi]^{-1/2} exp(-y_0^2/2)$.
The conditional probability $Q(x_0|y_0)$
 satisfies another FP equation, that can
be solved iteratively for small $\alpha/\nu$ \cite{bj}.
For large $x_0$ we have obtained the first three terms of
$p(x_0)=\int P(x_0,y_0)dy_0$ expanded as a {\em formal} power series:
$p(x_0)=\sum_{n=0}^{\infty}(\alpha/\nu)^n p_{n}(x_0)$.
The functions $p_0,p_1,p_2$ are combinations of
modified Bessel functions that have exponential tails
for large values of $x_0$ with
$p_0(x_0) = [\pi]^{-1} K_0(|x_0|)$, where $K_0$ is the modified
Bessel function of order zero. Note that the
exponential tail results from  a linear FP equation
due to a branch cut in the characteristic functions
in contrast to Ref.5 in which
a nonlinear FP equation is used with
simple poles in the characteristic function.
In the limit of very large
$\alpha/\nu$, $p(x_0)$ becomes Gaussian.
We compute the low-order moments and find
that the kurtosis is given by
${\bf K} =  3+\frac{6\nu}{\alpha+\nu}$; this interpolates
between 9 for small $\alpha/\nu$ and 3 for
large $\alpha/\nu$  corresponding to $p_0$ and
the Gaussian distribution respectively.
In Fig.4 we show the PDF of $x_0$ obtained
from numerical simulations for  $\alpha=\nu$ for which ${\bf K}=6$
corresponding to that of an exponential distribution.

    The form of the PDF depends not only on the parameter
$B$ but also on the nature of the velocity correlations.
In addition to exponential tails discussed above other non-Gaussian
behavior can occur in different models.
For example, the introduction of another variable $z$ using
$\dot{z}=-\mu z + x \eta_3$ in the toy model (Eq.3,4)
leads to a stretched exponential PDF for $z$ with an
exponent $2/3$ for appropriate $\alpha,\nu$ and $\mu$. A similar
generalization of the (stochastic) model A can be constructed.
We emphasize that the PDF
of the variable depends on the distribution of the
slowly varying amplitude of the
noise but not on the distribution of the noise itself.

We now consider the model in which the  velocity is
obtained from $\psi_n(i) = \eta_n(i)$,
where $\eta_n(i)$ is a Gaussian noise correlated
over a time $\tau_c$ with a correlation length $\xi_v$.
The PDFs of $\theta$ are Gaussian when $B$
is smaller than unity even for $\tau_c \approx \tau_d$.
This is true for both the discrete time and the
continuum model. However, for $B > \approx  1$
the PDFs are non-Gaussian and we study this case
using the continuum model {\em i.e.,} Eq.(1).
The resultant PDFs are shown in Fig.5 for different parameters.
(i) For $B>>1$ the distributions (Fig.5a,5b,5c)
clearly do not have exponential tails; however,
strikingly, we could fit the PDFs very well
with a modified Lorentzian with two
parameters ($\kappa, \delta$):
$p(x)= C/(1+\kappa x^2)^{1+\delta}$ where
$C$ is fixed by normalization. This function is a  variant of the
form proposed by Sinai and Yakhot\cite{sy}.
(ii) For $B \approx 1$ the scalar
PDFs appear to have exponential tails over a narrow range of parameters
(this case corresponds to Fig.5d).
However, the fit with a modified Lorentzian agrees over a wide range of
$B$ values including this narrow range.

  Since the convective noise term is not negligible we model
the $B > \approx 1$ regime  by a single variable with both an
additive ($\zeta_1$) and a multiplicative
($\zeta_2$) noise term:
\begin{equation}
\dot{x}=-\nu x +\zeta_1 + \zeta_2 x.
\end{equation}
When $\zeta_i$ are white noise with
$<\zeta_i(t)>=0$ and $<\zeta_i(t)\zeta_j(t')> =
2D_i\delta_{ij}\delta(t-t')$, a straightforward solution
of the Fokker-Planck equation yields exactly the
modified Lorentzian form for the PDF
with $\kappa=D_2/D_1$, $\delta=\nu/2D_2$ \cite{note}.
This provides another example of a simple toy problem that mimics
the PDFs of passive scalars.

We note that the exponential tail regime for passive scalars
in several experiments \cite{jw,gol} appears to correspond
to the $B < 1$ regime in our model where intermittent
velocity fluctuations lead to exponential tails.
Experimentally such amplitude variations may be
connected with coherent structures such as
plumes and turbulent bursts\cite{sg}.
A quite different extended dynamical system in which our
mechanism is operative is a Capacitive Josephson
Junction Array driven by external dc and ac currents;
numerical simulations in chaotic states indicate
exponential tails in the PDFs of local junction voltages
transverse to the external current\cite{bej}. This can be
ascribed to intermittent current fluctuations caused by
random vortex motion. However, establishing the occurrence of
such amplitude fluctuations starting
from the underlying equations, {\em e.g., } Navier-Stokes
equations with appropriate boundary conditions, remains a
challenge.

This work was supported by the U. S. Department
of Energy (Contract No. DE-F-G02-88ER13916A000). We acknowledge
computer time on the Cray-YMP provided by The Ohio
Supercomputer Center. CJ is grateful to Dr. Yu He for valuable
discussions; RB thanks Jayesh for helpful conversations.

\pagebreak

\section{Figure captions}

Fig.1: PDF for the normalized fluctuation
$X=\delta\theta/\sigma_{\theta}$ for Model A ({\em stochastic})
with $ L=48, \nu=0.1,\gamma=0.4,\beta=0.1, \tau_a \approx 10$.
The white noise terms are uniformly distributed between -1 and 1.
The variances are $\sigma_{\theta} \approx 0.034,\sigma_v \approx 0.21$
and the kurtosis is 4.22.

Fig.2: Data for Model B ({\em deterministic}).\\
a) PDF for  $v$ normalized by $\sigma_v$ has Gaussian tails.
b) PDFs for the normalized fluctuation $X$ are shown
 for two parameters. The dashed line is drawn
with the same slope as that of the tail. i) Upper curve
(shifted up by two decades) is for $\nu=0.1,\gamma=0.5,\beta=0.1$.
The variances are: $\sigma_{\theta} \approx 0.045,\sigma_v \approx 0.2$
and the kurtosis is 4.62. ii) Lower curve  is for
$\nu=0.05,\gamma=0.25,\beta=0.1$. The variances are
$\sigma_{\theta} \approx 0.031,\sigma_{v} \approx 0.1$
and the kurtosis is 4.12. Data are obtained using 5 million
points in intervals of 10 time steps
at a site in the middle of a 48 x 48 lattice.

Fig.3:  Correlation functions of the velocity used in Fig.2.
A semilog plot of $C_4(n)$ vs $n$ is shown. The straight lines have slopes
$\approx 1.3$ and $\approx 0.24$. The inset shows a
semilog plot of $C_2(n)$ vs $n$; the straight line has a
slope $0.69$. See text for discussion.
(The variance  $\sigma_v$ is set to unity.)

Fig.4: PDF for $x_0=x/\sigma_x$ in the toy model of
Equations (3) and (4) with $\alpha=\nu$. The noise variables
$\eta_1,\eta_2$ are uniformly distributed with $\sigma_1=\sigma_2=1$
and the measured $\sigma_x=0.5$.

Fig.5: PDFs for the passive scalar fluctuations with $B >\approx 1$
using the continuum model Equation (1).
The data are obtained with $L=48,\nu=0.1,\beta=0.1$
using the discretizations $\delta t=0.05,\delta l =0.5$ and
velocity correlations $\xi_v \approx \delta l$ and $\tau_c=2$.
The upper three curves (a,b,c) are shifted up from the
lowest curve (d) by 6,4 and 2 decades respectively.
The curves a,b,c,d correspond to
$\sigma_v=1.4 \gamma$ where $\gamma=(0.4,0.35,0.3,0.2)$
with kurtosis values $(15.6,6.3,4.1,3.5)$. The dashed lines are
modified Lorentzian (ML) fits for the data.


\begin{thebibliography}{99}
\bibitem{cetal} B. Castaing {\em et. al}, {\it J. of Fluid Mech., }
{\bf 204}, 1 (1989).

\bibitem{jw} Jayesh and Z. Warhaft, {\it Phys. Rev. Lett.},
{\bf 67}, 3503 (1991); {\em Physics of Fluids A.} {\bf 4}, 2292 (1992).

\bibitem{gol} J. P. Gollub, J. Clarke, M. Gharib, B. Lane,
and O. N. Mesquita, {\it Phys. Rev. Lett.}, {\bf 67} 3507 (1991).

\bibitem{cgh} B. Castaing, Y. Gagne, and E. Hopfinger, {\em Physica },
{\bf 46D}, 177 (1990). See also, M. Sano, X. Wu, and A. Libchaber,
{\it Phys. Rev. A.}, {\bf 40}, 6421 (1989).

\bibitem{pss} A. Pumir, B. Shraiman, and E. D. Siggia,
{\it Phys. Rev. Lett.}, {\bf 66},
 2985 (1991); M. Holzer and A. Pumir,
{\it Phys. Rev. A.}, {\bf 47}, 202 (1993).

\bibitem{sy} Ya. G. Sinai and V. Yakhot, {\it Phys. Rev. Lett.},
 {\bf 63}, 1962 (1989).

\bibitem{ycck} V. Yakhot, {\it Phys. Rev. Lett.},
{\bf 63}, 1965 (1989); H. Chen, S. Chen,
and R. H. Kraichnan, {\it Phys. Rev. Lett.}, {\bf 63}, 2657 (1989).

\bibitem{ching} E. S. C. Ching, {\it Phys. Rev. Lett.},
{\bf 70}, 283 (1993);S. B. Pope and E. S. C. Ching, Preprint.

\bibitem{ker} A. Kerstein, {\it J. of Fluid Mech., }
{\bf 231}, 361 (1991), also see
 for the direct simulations,  O. Metais and M. Lesieur,
{\it J. of Fluid Mech., } {\bf 239}, 157 (1992).

\bibitem{bound} One can set a bound for a given $\nu$
from the linear stability analysis. Also, for $\nu \geq 0.25$ the
uniform state is unstable.

\bibitem{risk} H. Risken, {\em Fokker-Planck Equation}
 (Springer Verlag 1989).

\bibitem{bj} R. Bhagavatula and C. Jayaprakash, in preparation.

\bibitem{note} Note that for a specific choice $D_1=\nu$
the distribution reduces to the PDF derived in Ref.6.

\bibitem{sg} T. H. Solomon and J. P. Gollub, {\it Phys. Rev. Lett.},
 {\bf 64}, 2382 (1990). S. Douady, Y. Couder, and M. E. Brachet,
{\it Phys. Rev. Lett.}, {\bf 67}, 983 (1991).

\bibitem{bej} R. Bhagavatula, C. Ebner, and C.Jayaprakash, to be
published.

\end{thebibliography}
\end{document}